# Evaluation of a new cavitation erosion metric based on fluid-solid energy transfer in channel flow simulations

G. M. Magnotti[1,*], M. Battistoni[2], K. Saha[3], S. Som[1]
[1]Argonne National Laboratory, Argonne, IL, USA
[2]Università degli Studi di Perugia, Perugia, Italy
[3]Bennett University, Greater Noida, India

**Abstract**
Although there have been extensive investigations characterizing cavitation phenomenon in fuel injectors, much is still unknown about the mechanisms driving cavitation-induced erosion, and how these complicated physics should be represented in a model. In lieu of computationally expensive fluid-structure interaction modeling, the Eulerian mixture modeling approach has been accepted as an efficient means of capturing cavitation phenomena. However, there remains a need to link the erosive potential of cloud collapse events with the subsequent material deformation and damage of neighboring surfaces. Even though several cavitation erosion indices have been proposed in the literature, no single metric has been identified as universally applicable across all injector-relevant conditions.

The objective of this work is to identify parameters that characterize the erosive potential of cavitation cloud collapse mechanisms that are likely to occur within injector orifices. While a commonly employed cavitation erosion metric, namely the maximum local pressure, was found to provide indications of potential sites for pitting and material rupture from single impact events, no additional information could be determined regarding the material erosion process. To improve representation of the incubation period within the cavitation erosion process, a new metric was derived based on cumulative energy absorbed by the solid material from repeated hydrodynamic impacts. Through evaluation of predicted cavitation cloud collapse events in a channel geometry against available experimental data, the stored energy metric yielded insight into the erosive potential of recorded impact events. The stored energy metric provided a means to accurately predict the influence of flow conditions on the incubation period before material erosion. Additionally, detailed analysis of cavitation cloud collapse events preceding impacts suggests that the cloud collapse mechanism governs the erosive potential of impacts and the resultant incubation period. Specifically, horseshoe cloud implosions were found to yield higher impact energies relative to spherical cloud collapse events.

**Introduction**
Cavitation-induced erosion has been studied using experimental [1]-[6] and computational modeling techniques [6]-[10] to gain insight into the physical mechanisms driving this process for a range of applications, from pump impellers to ship rudders. Cavitation erosion also has been noted in diesel injector hardware, namely within the needle seat region and along the injector needle, as well as the entrance to nozzle holes [6], [8]. A common feature among these vulnerable locations is the proximity to an area contraction, where local flow acceleration and pressure reduction promote cavitation. The erosive potential of the cavitation shedding processes in these regions have been found to be highly dependent on the local geometric features and fluid properties.

Ideally, an injector hardware designer would be able to utilize a computationally efficient design tool to accurately predict these cavitation and condensation events, and to inform improved designs that mitigate the severity of cavitation-induced erosion. In lieu of computationally expensive fluid-structure interaction modeling [11], the Eulerian mixture modeling approach has been accepted as a computationally efficient means of capturing cavitation phenomena when sufficient resolution is employed [10]. However, when employing such a framework, there remains a need to link condensation events with cavitation erosion potential and material fatigue.

Developing a predictive erosion index is a challenge due to the disparate timescales characterizing cavitation impacts and the gradual material fatigue and erosion process. Although several cavitation erosion indicators have been proposed in the literature [6]-[9], [12], no single metric has been identified as universally applicable across all injector geometries and injection conditions. The majority of cavitation erosion indices implemented in today's computational fluid dynamics (CFD) codes characterize the hydrodynamic conditions following critical cloud implosion events. For example, the maximum local pressure [8] provides an efficient means to represent the extrema of impact stresses on the neighboring walls. Although this metric can characterize the potential for plastic deformation and pitting from impacts in excess of the material yield stress, the maximum local pressure does not provide insight into the material response from repeated hydrodynamic impacts. In contrast, the mean depth penetration rate (MDPR) parameter [12] is one of the few metrics in the literature that provides a detailed treatment of the material response. The MDPR metric includes the influence of strain hardening, and elastic and plastic deformation on the lifetime of the material using bulk properties obtained from tensile tests [12], namely the material yield stress and strain. However, the main limitation of this erosion index is the simplified representation of the fluid mechanics. More specifically, the influence of cloud collapse on the neighboring walls is represented using a single set of impact stress and strain values. The ability of MDPR to capture the effect of repeated cavitation impacts is uncertain due to the large discrepancy in strain rates used to characterize the bulk material properties and those estimated from cavitation impacts ($10^3 - 10^4$ s$^{-1}$ [13]). Therefore, there exists an opportunity to formulate an improved erosion metric that can bridge the existing gap in how repeated cavitation cloud implosions are represented and linked to the material erosion process.

In order to progress towards identifying a predictive and robust cavitation erosion indicator, we focus our efforts on modeling the flow of pressurized fuel through a simplified channel geometry with an inlet diameter of approximately 300 μm [1], which serves as a geometric analogue to an injector nozzle. We first assess the ability of the maximum predicted pressure at the throttle surface boundaries to characterize the propensity for cavitation erosion, and evaluate its link to the observed cavitation structures. After comparison with available erosion testing data quantifying first erosion sites and incubation timescales [1], a new metric is proposed that characterizes critical locations where repeated condensation events are likely to occur and lead to material fatigue and erosion.

**Simulated Conditions for PREVERO Channel K**

Using best practices from previous computational investigations [10], [14]-[17], CONVERGE [18] is utilized to model cavitation within the PREVERO Channel "K" geometry from the experimental work of Skoda et al. [1] for pressure drops ranging from 150 to 265 bar. Because the experimental data set includes characterization of erosion events within this geometry, this case serves as the starting point to validate the current cavitation modeling configuration, and study the predicted cavitation cloud structures with erosive potential. The key parameters describing the experimental conditions explored in this work are defined in Table 1. Diesel fuel is represented as a barotropic fluid using fluid properties for n-heptane and a reference density of 833.6 kg/m$^3$ at a reference pressure of 30 MPa [19]. A trace amount of non-condensable gas is assumed to be present in the liquid fuel, as represented with an N$_2$ mass fraction of 2.0e-5 [10]. An illustration of the Channel K geometry and simulated domain is shown in Figure 1. Key features of the channel geometry include a constant diameter of 303 μm, channel length of 994 μm and an inlet radius of curvature of 40 μm. It is important to note that in the experiments of Skoda et al. [1], the channel was constructed from aluminum in order to accelerate the expected incubation period before material erosion.

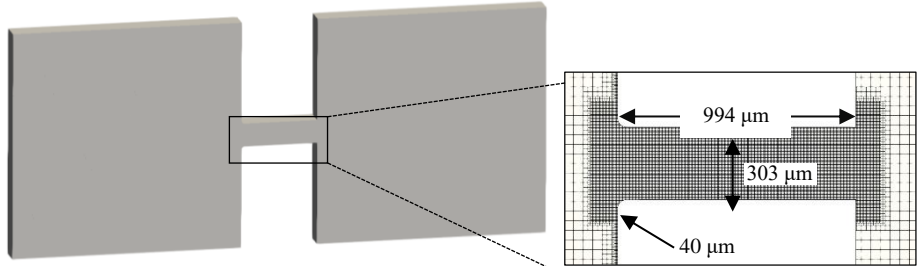

**Figure 1.** Illustration of simulated domain, with details of fixed embedding in channel geometry.

**Table 1.** Modeled conditions within the PREVERO Channel "K" Geometry.

| Liquid Fuel | Fuel Temperature [K] | Upstream Reservoir Pressure [bar] | Downstream Reservoir Pressure [bar] | Pressure Drop Across Throttle [bar] | Nitrogen Mass Fraction |
|---|---|---|---|---|---|
| n-heptane | 327 | 300 | 35 - 150 | 150 - 265 | 2.0e-05 |

**Model Formulation**

The cavitating flow within the channel is modeled as a three-component two-phase fluid system, comprised of liquid and vapor fuel and non-condensable gas, represented using N$_2$. Within the homogeneous mixture modeling approach, thermal and mechanical equilibrium are assumed, whereby all components are assumed to have the same pressure, temperature and velocity within a given computational cell. Using a pseudo-density concept, the mixture density, $\rho$, is defined using a volume-weighted average of the components,

$$\rho = \sum_{i=1}^{3} \alpha_i \rho_i \tag{1}$$

where $\alpha_i$ and $\rho_i$ are the void fraction and density of the $i^{th}$-component, respectively. For gaseous phases, the density is determined using the Redlich-Kwong equation of state, whereas the liquid phase density is determined using a barotropic relation for compressible liquids. Unlike standard volume of fluid (VOF) methods, $\alpha$ is not directly transported. Rather, $\alpha$ is determined via the species mass fraction, $Y_i$, which is solved for in the species transport equations,

$$\frac{\partial \rho Y_i}{\partial t} + \nabla \cdot \rho Y_i \vec{v} = \nabla \cdot (\rho D_i \nabla Y_i) + S_i \tag{2}$$

where $\vec{v}$ and $D_i$ are the velocity and diffusivity of the $i^{th}$-component, respectively, and $S_i$ is the source term due to mass transfer (i.e., cavitation and condensation).

Cavitation and condensation are represented using the homogeneous relaxation model (HRM) [20], where the rate at which the instantaneous quality of fuel, $x$, approaches its equilibrium value, $\bar{x}$, is defined with the following relation,

$$\frac{Dx}{Dt} = \frac{\bar{x} - x}{\theta} \tag{3}$$

$$\theta = \theta_0 \alpha^{-0.54} \psi^{-1.76} \tag{4}$$

where $\theta$ is the relaxation time scale. For cavitation, the phase change timescale is set equal to $\theta$, where $\theta_0$ is a coefficient set to 3.84e-7, and $\psi$ is the non-dimensional pressure ratio ($\psi = \frac{p_{sat} - p}{p_{crit} - p_{sat}}$). Although the presence of non-condensable gas is included in this model, no additional models are included to represent the adsorption and absorption of $N_2$. As a result, the mass transfer source term in the $N_2$ species transport equation is set to zero. To represent the turbulent flow structures within the throttle, the large eddy simulation (LES) dynamic structure model was employed [21]. In agreement with previous studies [14]-[15], a grid convergence study revealed that a minimum cell size of 5.0 μm was sufficient to capture cavitation development in the channel.

**Cavitation Erosion Metrics**

Several metrics have been proposed in the literature to characterize the cavitation cloud implosion events and their contribution to material erosion. As previously discussed, existing metrics are incomplete in their ability to quantify the hydrodynamic impulse loads acting on neighboring surfaces and link the predicted impacts with the progressive material fatigue and rupture process. To highlight the need for an improved metric, an evaluation of the predicted erosive potential is conducted using a popularly utilized metric, namely the maximum local pressure [7]. The maximum local pressure is calculated using a user-defined function (UDF) to track the maximum predicted pressure along the surface of the channel boundary. Through comparison with the material specifications of aluminum for the yield stress, this metric offers an opportunity to identify potential locations for single impact events leading to pitting. However, these single impact events do not encapsulate the series of hydrodynamic impacts within the incubation period that lead to material fatigue and failure. As a result, parameters like the local maximum pressure are not able to provide insight into the gradual cavitation erosion process, or its dependence on local flow conditions and geometric parameters.

In order to link the effect of repeated impacts, of varying stress and strain rates, with the eventual material fatigue and rupture, a new metric is proposed. In the development of this parameter, several assumptions and approximations are made to characterize the erosive potential of cloud implosions and simplify the representation of material deformation. The impact stress at the wall is quantified using the predicted pressure at the wall boundary. The properties of aluminum are modeled as a bilinear material, where an elastic response occurs for stresses lower than the yield stress and plastic deformation results from stresses larger than the yield stress, but smaller than the ultimate stress. As such, for impact loads smaller than the yield stress, the material is assumed to respond elastically, where the material returns to its initial stress condition, and no damage occurs. In this work, this limit is imposed at 60 MPa, based on reported specifications for aluminum 6082-O under static loading conditions [22]. For impact stresses larger than the yield stress, a portion of impact energy is absorbed and stored in the material. In the current implementation of the cavitation erosion metric, no additional treatment is provided for work hardening or other sub-surface material effects. Therefore, the absorbed impacts are treated as stored energy within the superficial layer of the material.

Using these assumptions, a new metric characterizing the progress towards material erosion is derived based on an energy analysis of a control volume at the fluid-solid interface, as depicted in Figure 2. Because the control volume is drawn around the fluid-solid interface, no assumptions are made about the impact energy source. Instead, a given cloud collapse event is represented by its impact energy, $E_{impact}$, that is transferred to the wall based on local fluid properties. $E_{impact,i}$, for a given impact event $i$, is defined by considering the energy of the pressure wave acting on the wall [23],

$$E_{impact,i} = \frac{\mathcal{A}}{\rho_l c_l} \int_0^\tau p^2(t) dt \tag{5}$$

where a pressure wave of magnitude $p$ propagates through a medium, characterized by its acoustic impedance, defined as the product of its density $\rho_l$ and speed of sound $c_l$, and acts on the surface of area $\mathcal{A}$ over a duration of time $\tau$. The time interval for a given impact is defined by the time instances when the pressure acting on the wall exceeds the yield stress of the wall material. In the current implementation, the impacted area $\mathcal{A}$ is based on the size of the wall-neighboring cells, which is equal to the minimum cell size of 5 μm. The fluid near the wall is assumed to be pure liquid based on analysis of predicted cloud implosion events, which occur downstream of the vapor cloud cavity. Even if small concentrations of bubbles are present near the wall, the grid resolution employed in the mixture model approach likely limits the contribution of these bubbles to the mixture speed of sound. As a result, the assumption of liquid as the medium through which the pressure wave acts upon the wall is a reasonable one.

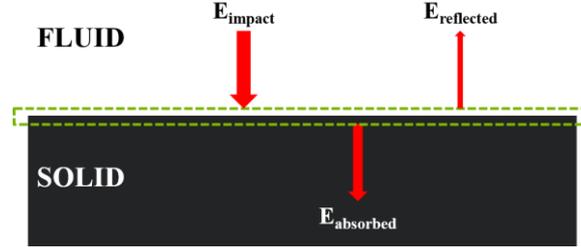

**Figure 2.** Schematic of energy balance considered in the derivation of the new cavitation erosion metric, which considers the transfer and storage of energy in the solid material due to repeated impacts from cavitation cloud collapse events.

Based on the impact stress and relative acoustic impedances of the fluid and solid, a portion of $E_{impact}$ may be reflected back into the fluid, as denoted by the outflow of $E_{reflected}$ from the control volume. For impact stresses less than the yield stress of aluminum (60 MPa), the solid is assumed to behave elastically, where none of the impact energy is absorbed by the material. As a result, for impact stresses less than the yield stress, 100% of the impact energy is assumed to be reflected. For impact stresses greater than the yield stress, calculation of the ratio of the acoustic impedances of n-heptane [24] and aluminum [25] reveals that less than 5% of the impact stress would be reflected due to dampening effects from recoil of the wall [26]. Based on this information, the aluminum walls of the channel are treated as perfectly rigid, and the dampening effect of the aluminum-heptane system is neglected. As a result, for a given impact $i$ characterized by a pressure above the yield stress, all of the impact energy is assumed to be absorbed by the material:

$$E_{absorbed,i} = E_{impact,i} \quad (6)$$

Based on this approach, the solid material can be likened to a high-pass filter, where stress loadings below the yield stress are reflected, whereas those above the yield stress are absorbed by the material.

To represent the progressive damage to the material from repeated impacts, a cumulative stored energy by the material is calculated. The total impact energy stored in the material after a given number of impacts, $N$, is determined by combining Equations (5) and (6) and integrating over the discrete impact events, as shown in the following relation,

$$E_{stored}(N) = \sum_{i=1}^{N} E_{impact,i} = \sum_{i=1}^{N} \frac{\mathcal{A}}{\rho_l c_l} \int_0^\tau p^2(t) dt. \quad (7)$$

In this approach, material failure would then be predicted when $E_{stored}$ exceeds a critical threshold. However, determination of the critical energy threshold for material rupture is not a straightforward calculation. Within the incubation period, a multitude of cavitation cloud collapse events with varying strain rates can be expected to occur [13]. As a result, a set of critical stress and strain parameters, conventionally measured under a constant strain rate condition, would not be likely to characterize the resultant conditions at failure [13]. In the current stage of this work, a critical energy is not yet defined, but is deemed as an important parameter to be characterized in future investigations. In its current form, $E_{stored}$ is used as a qualitative measure of cavitation erosion. However, based on its ability to capture the effect of repeated impacts on the material state and current progress within the incubation period, the newly derived $E_{stored}$ provides an improved characterization of cavitation erosion over existing metrics in the literature.

**Results and Discussion**

To validate the ability of the modeling approach to capture the thermofluidic conditions leading to cavitation erosion, predictions of mass flow rate at the channel exit are first compared with the experimental data from Skoda et al. [1] for a range of pressure drop conditions from 150 to 265 bar, as shown in Figure 3. Overall, excellent agreement is achieved between the predicted and measured mass flow rates across the wide range of pressure drops evaluated in this study. Additionally, the critical pressure drop condition of 200 bar, where choking is observed and the mass flow rate plateaus with increasing pressure drop, is also well matched. This validation exercise suggests that the fluid properties and treatment of the multiphase flow in this modeling approach well represent the experimental flow conditions at the channel exit.

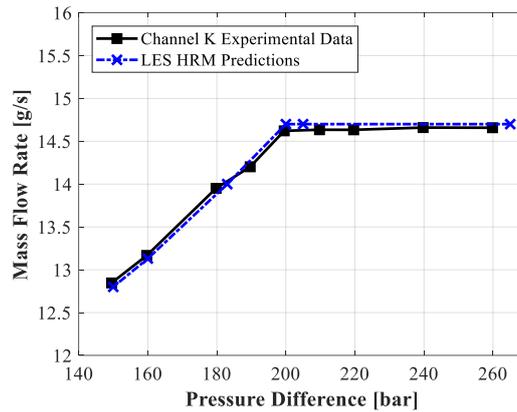

**Figure 3.** Comparison of measured and predicted mass flow rates for a range of pressure drop conditions across the Channel "K" geometry [1].

In order to validate the predicted cavitation cloud structures, model predictions of the time-averaged line-of-sight integrated total void fraction distribution are then compared to the extent of cavitation indicated in the experimental images of Skoda and co-workers [1], as shown in Figure 4. Experimental images are shown in Figure 4(a)-(b) for pressure drop conditions of 183 bar ("OP2") and 247 bar ("OP3"), respectively. The extent of cavitation in the experiments was characterized with the cloud collapse location (CCL), defined as the location with a cavitation probability of 30%. In comparison to the modeling results, CCL corresponds to a time-averaged void fraction of 0.30. Although the predicted and measured mass flow rates were well matched at the OP2 and OP3 conditions, as previously shown in Figure 3, the time-averaged CCL's were found to be underpredicted at the specified experimental conditions. After extensive exploration of the sensitivity of predicted CCL to physical and numerical parameters in the models (not shown in this paper for the sake of brevity), the outlet reservoir pressure was modified to yield improved agreement with the experimentally indicated CCL. The modified pressure drop conditions, $\varDelta P^*$, are annotated in Figure 4(c) and (d), and are within 12% and 7% of the experimentally specified pressure drops for the OP2 and OP3 conditions, respectively. This modification of the simulated conditions is supported by the computational study conducted by Skoda et al. [1], who also noted the inability to match the experimental results at the specified conditions. The consistent boundary condition modification required by two different codes to match the experimental data suggests that the source of discrepancy is likely due to uncertainty in the experimental conditions, as opposed to numerical treatment of the cavitation process. Although the $\varDelta P^*$ conditions allow for more straightforward comparison of the predicted and measured cavitation cloud structures, and ultimately cavitation erosion sites, care must be taken when interpreting the predicted cavitation erosion events. In the simulated OP3 condition, the increased pressure drop is not expected to influence the predicted frequency or intensity of cloud collapse events due to the choked flow condition, which results in a constant average flow velocity at the channel exit for pressure drops larger than 200 bar. However, for the simulated OP2 condition, the increased pressure drop will induce larger flow velocities than would be expected at the experimental condition, and thus may result in more rapid cloud collapses and higher predicted impact pressures than may have occurred in the experiment. These effects should therefore be taken into consideration when evaluating the predicted cavitation erosion events.

Using the $\varDelta P^*$ conditions to match the measured CCL, the predicted severity of cavitation erosion can now be compared with the experimental data from Skoda et al. [1]. Experimental images from the cavitation erosion study are shown in Figure 5(a) and (b) for the OP2 and OP3 conditions, respectively, at the time instant marking the end of the incubation period, *T*. At the OP2 condition, an incubation period of 45 minutes is recorded before the first occurrence of material erosion, whereas a larger incubation period of 60 minutes is noted to occur for the OP3 condition. These results suggest a higher cavitation erosion intensity for the OP2 condition relative to the OP3 condition. The location of material erosion is also shown to have a dependence on the pressure drop condition. Under the OP2 condition, the first site of material erosion occurs at approximately 70% of the channel length, whereas material erosion is not observed until the channel exit at the OP3 condition.

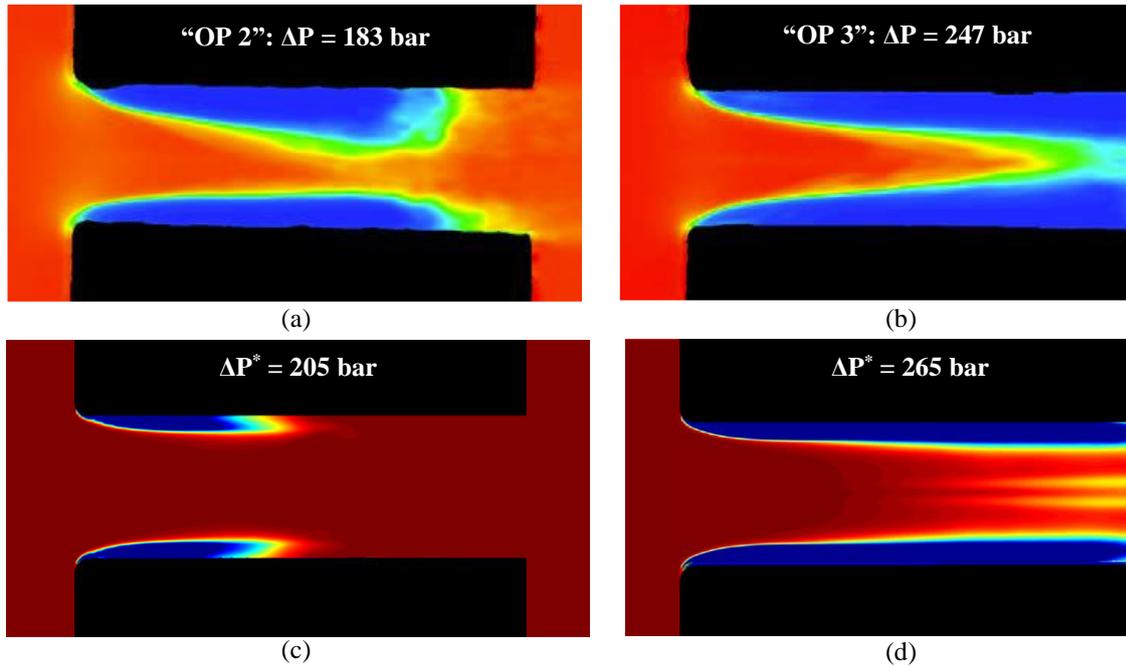

**Figure 4.** Comparison of total void fraction between (a-b) experimental images from Skoda et al. [1] and (c-d) time-averaged line-of-sight integrated HRM model predictions at modified back-pressure conditions to match the experimentally measured cloud collapse location.

Using the implemented computational metrics as indicators of cavitation erosion, namely the maximum local pressure and $E_{stored}$, the ability of the model to capture material erosion can be evaluated. Additionally, the utility of each of these metrics in characterizing aspects of the cavitation erosion process can be ascertained. Predictions of cavitation erosion with the maximum local pressure metric are shown in Figure 5(c) and (d) for the OP2 and OP3 conditions. Results for the OP2 condition indicate impact stresses in excess of 30 MPa at locations between 34-60% of the total channel length, which is in acceptable agreement with the experimentally indicated location at approximately 70% of the channel length. Within the 400 μs evaluation window, or approximately 80 flow-through times, several repeated impacts of similar magnitude are predicted in this region. At the OP3 condition, a different maximum pressure distribution is observed. Within the same 400 μs period of time, 80% of the entire channel length is predicted to experience local pressures in excess of 30 MPa. Localized impacts are clearly visible at the extents of this region. While the maximum local pressure metric provides an indication of potential sites for pitting and material rupture from single impact events, no additional information can be determined regarding the incubation period or critical locations where material erosion is likely.

To gain further insight into the predicted cavitation erosion process, the amount of stored energy in the material due to repeated impacts after a 400 μs window is shown in Figure 5(e) and (f) for the OP2 and OP3 conditions, respectively. At the OP2 condition, several of the impacts indicated by the maximum local pressure metric in Figure 5(c) are filtered out, and only impacts with stresses in excess of 60 MPa are shown. The average stored energy from these impacts is predicted to be approximately 7 nJ. For the OP3 condition, several high energy impacts are revealed at locations upstream of the channel exit. The stored energy at these locations were found to be from single impact events during transient recession of the cavitation cloud. Because the impact pressures at these locations as indicated in Figure 5(d) are less than the ultimate stress of aluminum 6082-O of 300 MPa [22], these high energy impacts are not likely to result in material rupture. Within the 400 μs window, no repeated impacts were predicted at these locations, suggesting that these locations are not probable cavitation erosion sites. However, near the channel exit, several repeated impacts of similar energy are observed. The average stored energy from these impacts is predicted to be approximately 5 nJ. Comparison of the predicted stored energy distributions for the OP2 condition and at the channel exit for the OP3 condition provides a qualitative assessment of relative incubation period before material rupture. The larger average stored energy for the OP2 condition in comparison to the OP3 condition suggests that the incubation period would be shorter under the OP2 condition. This result is consistent with the experimental findings, where the incubation period for the OP3 condition was found to be relatively longer. These findings using the newly developed cavitation erosion metric highlight its utility in quantifying both the energy of single impact events, as well as the influence of repeated impacts on the incubation period before material rupture. Future work will evaluate longer simulated times in order to quantify the rate of energy storage, and how this information can be used to extrapolate behavior dictating the incubation period, which occurs over a much longer period of time than can be feasibly simulated [13].

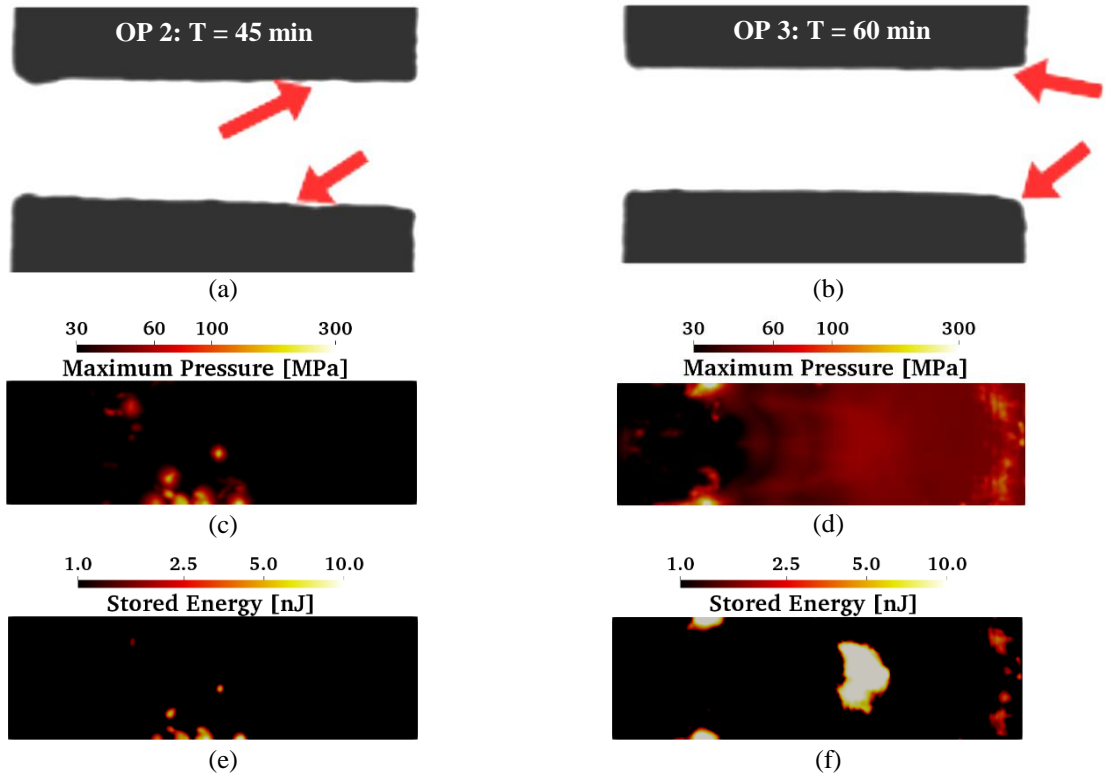

**Figure 5.** Comparison of erosion patterns between (a-b) experimental images from Skoda et al. [1] and HRM model predictions of (c-d) local maximum pressure and (e-f) accumulated stored energy at the top channel wall, which are employed as indicators of cavitation erosion potential.

To uncover the difference in physical mechanisms driving cavitation erosion severity for the OP2 and OP3 conditions, the predicted cavitation shedding and condensation events preceding critical impact events are evaluated in detail. For the OP2 condition, a visualization of the predicted cavitation cloud development prior to an impact in excess of 60MPa is shown in Figure 6. The cavitation cloud structure, shown in pale yellow, is visualized using iso-contours of 15% total void fraction. Downstream of the cavitation cavity formed at the channel inlet, a cavitation cloud that was shed in a previous time instant has evolved into a horseshoe-shaped cloud. Based on the work from Dular and Petkovšek [2], the physical mechanism governing cavitation erosion is strongly linked to the generation and development of vortices in the cavitation shedding process. As a result, vortical structures were visualized using a Q-criterion iso-contour of $4E12s^{-2}$, and filtered using a built-in connectivity algorithm in ParaView [27]. The horseshoe vortex shown in Figure 6 is colored by the velocity normal to the wall, where negative values indicate flows directed towards the lower wall. The simultaneous visualization of cavitation cloud and vortex structures clearly shows that the horseshoe cloud implosion mechanism is ultimately responsible for the predicted surface damage. For the OP3 condition, within the evaluation time window, damage at the exit of the channel was not observed to be due to horseshoe cloud implosion; instead, predicted impacts were found to be governed by spherical cavitation cloud collapse. During intermittent periods when the cavitation cloud receded into the channel, the wall became exposed to pressure waves emitted from the collapse of condensing spherical clouds. Consistent with the findings from Dular and Petkovšek [2], damage from horseshoe cloud implosion was found to have a higher cavitation erosion potential than from spherical cloud collapse, as evidenced by the larger prediction of stored energy at the OP2 condition in comparison to that from the OP3 condition at the channel exit. Future work will evaluate large samples of critical impact events to gather statistics on predicted cavitation erosion mechanisms, and the frequency, duration, stress and energy of the associated impact events.

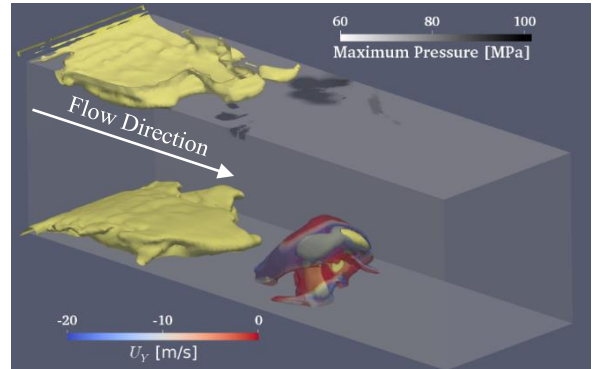

**Figure 6.** Visualization of cavitation cloud development preceding a high erosive potential impact event under the OP2 condition. Cavitation clouds are visualized with an iso-contour of void fraction set to 0.15, while the horseshoe cloud vortex is visualized using an iso-contour of the Q-criterion set to 4E12 s$^{-2}$.

**Summary and Conclusions**

To improve the link between predicted cavitation cloud collapse events and the incubation period leading to material rupture and erosion, a new computational metric was derived based on the fluid-solid energy transfer from impact events. Through comparison of predicted cavitation development and cloud collapse events in a channel geometry under two different pressure drop conditions with available experimental data, the stored energy metric revealed the following findings:

- While the maximum local pressure metric provides an indication of potential sites for pitting and material rupture for single impact events, no additional information could be determined regarding the incubation period or critical locations where material erosion is likely.
- The stored energy metric provides an indication of the influence of flow conditions on the incubation period before material erosion. These findings using the newly developed cavitation erosion metric highlight its utility in quantifying both the energy of single impact events, as well as the influence of repeated impacts on the incubation period before material rupture.
- The shorter incubation period indicated by the stored energy metric under the lower pressure drop condition was found to be linked to the different predicted cavitation erosion mechanisms. The horseshoe cloud implosion mechanism was found to have a higher cavitation erosion potential than the spherical cloud collapse, as evidenced by the larger prediction of stored energy.

To guide the continued development of the energy storage parameter as a measure of cavitation erosion, future investigations will focus on longer simulated times in order to quantify the rate of energy storage, and how this information can be used to extrapolate behavior dictating the incubation period, which occurs over a much longer period of time than can be feasibly simulated. Statistics will also be gathered to characterize the frequency, duration, stress and energy of predicted cavitation cloud collapse mechanisms that lead to critical impact events.

**Acknowledgements**

The submitted manuscript has been created by UChicago Argonne, LLC, Operator of Argonne National Laboratory (Argonne). Argonne, a U.S. Department of Energy Office (DOE) of Science laboratory, is operated under Contract No. DE-AC02-06CH11357. The U.S. Government retains for itself, and others acting on its be-half, a paid-up nonexclusive, irrevocable worldwide license in said article to reproduce, prepare derivative works, distribute copies to the public, and perform publicly and display publicly, by or on behalf of the Government.

Argonne National Laboratory's work was supported by the U.S Department of Energy under contract DE-AC02-06CH11357. The authors gratefully acknowledge the computing resources provided on Blues, a high-performance computing cluster operated by the Laboratory Computing Resource Center at Argonne National Laboratory, and Convergent Science Inc., for providing the CONVERGE CFD software licenses.